\documentstyle[12pt]{article}
\begin{document}
\baselineskip=0.7cm
\newcommand{\EQ}{\begin{equation}}
\newcommand{\EN}{\end{equation}}
\newcommand{\EQA}{\begin{eqnarray}}
\newcommand{\EQN}{\end{eqnarray}}
\newcommand{\e}{{\rm e}}
\newcommand{\Sp}{{\rm Sp}}
\newcommand{\Tr}{{\rm Tr}}
\makeatletter
\def\section{\@startsection{section}{1}{\z@}{-3.5ex plus -1ex minus
 -.2ex}{2.3ex plus .2ex}{\large}}
\def\subsection{\@startsection{subsection}{2}{\z@}{-3.25ex plus -1ex minus
 -.2ex}{1.5ex plus .2ex}{\normalsize\it}}
\def\appendix{
\par
\setcounter{section}{0}
\setcounter{subsection}{0}
\def\thesection{\Alph{section}}}
\makeatother
\def\thefootnote{\fnsymbol{footnote}}

\begin{flushright}
BROWN-HET-1135 \\
UT-KOMABA/98-18\\
August 1998
\end{flushright}

\begin{center}
\Large
Quantum Metamorphosis of Conformal Transformation 
\\
in D3-Brane Yang-Mills Theory

\vspace{1cm}
\normalsize
{\sc Antal Jevicki}
\footnote{
{\tt antal@het.brown.edu}
}

\vspace{0.3cm}
{\it Department of Physics, Brown University\\
Providence, RI 02912
}

\vspace{0.5cm}
{\sc Yoichi Kazama}
\footnote{
{\tt kazama@hep1.c.u-tokyo.ac.jp}
}
 \quad and \quad {\sc Tamiaki Yoneya}
\footnote{
{\tt tam@hep1.c.u-tokyo.ac.jp}
}

\vspace{0.3cm}
{\it Institute of Physics, University of Tokyo\\
Komaba, Meguro-ku, 153 Tokyo}

\vspace{1cm}
Abstract
\end{center}

We show how the linear special conformal transformation 
in four-dimensional $N=4$ super Yang-Mills theory 
is metamorphosed into the nonlinear and 
field-dependent transformation for the collective 
coordinates of Dirichlet 3-branes, which 
agrees with the transformation law for the 
space-time coordinates in the anti-de Sitter (AdS) 
space-time. Our result 
provides a new and strong support for the 
conjectured relation  between AdS$_5 \times S^5$
 supergravity  and super conformal Yang-Mills theory 
(SYM). 
Furthermore, our work sheds elucidating light on the 
nature of the AdS/SYM correspondence.

\newpage

One of the most interesting recent outcomes 
from studies of various duality relations in superstring 
theories is the correspondence between 
the large $N$ super Yang-Mills theories 
describing the low-energy dynamics of Dirichlet branes and 
supergravities in the background of 
anti de-Sitter space-times. Based on some 
earlier results \cite{klebanov}, the precise criteria for 
the validity of such correspondence are 
discussed in \cite{maldacena} and the concrete 
formulation of the relations between correlation functions 
on both sides are proposed 
in \cite{gubser-klebanov-polyakov} and 
\cite{witten1}.  From the viewpoint of 
string theory,   the correspondence 
may be regarded as a special version of old 
$s$-$t$ duality which relates open strings 
in $s$ channel and closed strings in $t$ channel. 
In connection with this view, 
it has been emphasized that, at the heart 
of this remarkable relation, there is an 
underlying duality between the short and large 
distances on each side, 
the `space-time uncertainty relation' in the 
language of \cite{jevicki-yoneya}\cite{li-yoneya} or `ultraviolet-infrared 
relation' using the terminology of \cite{witten-susskind}. 
 
As far as we can see, however,  the basis for the 
correspondence is yet phenomenological in its nature, and 
no logical or 
`intrinsic' explanation {\it within} the framework of 
super Yang-Mills theory  has been known, 
except for some analogy with lower dimensional 
examples such as those between the 
three-dimensional Chern-Simon theories in the bulk 
and the corresponding two-dimensional CFTs at the boundary.   
Very recently, some works 
\cite{douglas-taylor}\cite{berkooz}\cite{banksetal}\cite{balasetal} 
trying to fill this gap appeared. In the present letter, we 
provide another approach aiming toward such a goal. 

One of the handles in pursuing such an explanation   
is the (super) conformal symmetry on both sides. 
As is well known, the four-dimensional 
conformal group of 4D Yang-Mills theory is 
isomorphic to the isometry group of the 
five-dimensional anti de Sitter space-time, AdS$_5$.  
In the coordinate frame most appropriate 
for making comparison with the standard formulation 
of Yang-Mills theory, the metric 
on the supergravity side, AdS${_5}\times S^5$, is 
\begin{equation}
ds^2 = \alpha'\Bigl(
{R^2 \over U^2}(dU^2
+ U^2d\Omega_5^2) + {U^2\over R^2}dx^2_{4} 
\Bigr)  , 
\label{ads5metric}
\end{equation}
where and throughout the present paper 
 we use the same conventions as reference \cite{maldacena}. 
Thus $U$ is the radial 
coordinate measured in the energy unit, 
$U=r/\alpha'$. 
The throat radius of the AdS space-time in the dimesionless 
unit is  
$
R = (2g_{{\rm YM}}^2N)^{1/4}, 
$
and the Yang-Mills coupling is related to the standard 
string coupling $g_s=e^{\phi}$ by 
$g_{{\rm YM}}^2=2\pi g_s$. The special conformal transformation 
for the longitudinal four dimensional coordinates 
$x^a \, \, (a=0, 1, 2, 3)$ and the radial coordinate $U$ 
as a part of the isometry of this metric are 
\begin{eqnarray}
\delta_K x^a &=&  -2\epsilon\cdot x\, x^a +
\epsilon^a x^2 +\epsilon^a {R^4\over U^2}  ,
\label{specialadsx}\\ 
\delta_K\,  U &=& 2\epsilon\cdot x \, U .
\label{specialadsu}
\end{eqnarray}
In the usual interpretation of the Yang-Mills theory 
as the boundary field theory corresponding to 
supergravity in the bulk, the ordinary 
transformation law on the Yang-Mills side is  
identified with these transformations in 
the limit $U\rightarrow \infty$ of 
(\ref{specialadsx}) and (\ref{specialadsu}).  
This is certainly a consistent interpretation. 

On the other hand, from 
the viewpoint of effective world-volume theory 
for D3-brane, the Higgs fields of the $N=4$ 
super Yang-Mills theory must play the role of 
the collective coordinates which are transversal to the 
D3-brane, and hence correspond to the 
directions described by the radial coordinate $U$ 
together with the angle coordinates describing $S^5$ in 
the bulk theory.  
The above metric should therefore be detected in the dynamics 
of these coordinates representing a 
probe D3-brane in the presence of the 
background 
corresponding to the heavy source described by 
a large number of coincident 
D3-branes at rest at the origin.  From this point of view, 
the Yang-Mills theory as a whole 
{\it cannot} be regarded as living on the 
boundary (or anywhere)   
of the AdS space, since the above 
interpretation  crucially depends on the choice of 
D-brane configuration as the background of the 
Yang-Mills theory.  Clearly, 
the question "Where are the branes?" 
can meaningfully be asked only after a choice is made for the 
background in Yang-Mills theory. 

One way of approaching to this picture 
 is of course to try to 
compute the effective action for the probe D3-brane from the 
Yang-Mills side 
\cite{douglas-taylor}, just as we do for D-particles 
\cite{becke-becker-polchinski-tseytlin}\cite{okawa-yoneya} 
in Matrix theory.  In the following, we take a 
different approach. Namely, we try to derive the 
transformation law in the bulk theory directly from 
Yang-Mills theory.   In other words, we shall 
clarify how such a field-dependent transformation 
law can emerge from the ordinary linear 
transformation law. 

The classical Yang-Mills action in our convention is 
\begin{equation}
S_{d3}=-\int d^4 x \, 
{1\over 4g^2_{{\rm YM}}}\, {\rm Tr}
\Bigl(F_{ab}F^{ab} + 
{2\over (2\pi\alpha')^2}D_a X^{\mu}D^a X_{\mu}
+ {1\over (2\pi\alpha')^4}[X^{\mu}, X^{\nu}]^2 
\Bigr) + \cdots, 
\end{equation}
where we suppressed the fermionic part. Here 
the space-time indices  $\mu, \nu$ for the Higgs fields $X^{\mu}$ run through 
the transverse directions from 4 to 9 and 
the world-volume coordinates are identified with the 
space-time coordinates in the longitudinal directions 
$a=0, \ldots, 3$ assuming the static gauge 
for the parametrization in flat world volume. 
The action is invariant under the 
ordinary special conformal transformation 
generated by 
\begin{eqnarray}
\delta_K A_a (x)
&=&(\delta_K x^b) \partial_b A_a(x) 
-2\epsilon\cdot x A_a(x) 
+2x_a \epsilon\cdot A(x) - 2\epsilon_a x\cdot A(x) , \\
\delta_K X^{\mu}(x) 
&=&(\delta_K x^b)\partial_b X^{\mu}(x) 
-2\epsilon\cdot x 
X^{\mu}(x) .
\end{eqnarray}
The effective dynamics of D3-branes is described by the 
diagonal matrix elements of the Higgs fields. 
If the distance between 
the source and the probe is first assumed to be large, the 
energy-scale of the off-diagonal part is large or 
the length scale in the world volume 
is small, and it is appropriate to 
integrate over the off-diagonal part,  keeping fixed the 
low-energy (or large-distance) dynamics of the 
diagonal part.  In order to carry this out, we have to fix the 
gauge for the off-diagonal part. 
The most convenient is the usual background-field 
gauge, assuming the diagonal part $B$ of the fields as the background fields. 
Namely, the gauge function is 
$$F= \partial_a A^a -i{1\over 
(2\pi\alpha')^2}[B_{\mu}, Y^{\mu}] ,$$
where we have denoted the off-diagonal part of the 
Higgs fields by $Y^{\mu}$, i.e. $ X^{\mu}=B^{\mu}+Y^{\mu}$. 

Let us now  
consider the effect of the special conformal 
transformation on the gauge function $F$. 
We find 
\begin{equation}
\delta_K F =
(\delta_K x^a)\partial_a F 
-2\epsilon\cdot x F +4\epsilon\cdot A .
\end{equation}
Thus the gauge condition cannot be invariant under the 
special conformal transformation and, hence, 
we have to perform  a field dependent gauge 
transformation to recover the original gauge condition. 
The required gauge parameter is 
\begin{equation}
\Lambda = 4\triangle_B^{-1} \epsilon\cdot A  ,
\end{equation} 
where 
$\triangle_B$ is defined by $$\triangle_B \Lambda=-D^a\partial_a + {1\over (2\pi\alpha')^2}
[B_a+ Y_a, [B_a, \Lambda]] , $$ which of course is the 
kinetic operator for the ghost action 
$\int d^4 x \, {\rm Tr}\Bigl(
\overline{C}\triangle_B C\Bigr)
$. 
Thus the special conformal transformation for the 
Higgs fields is modified as 
\begin{equation}
\tilde{\delta}_K X= \delta_K X +
4i[\triangle_B^{-1}\epsilon \cdot A, X] .
\end{equation}
A similar modification of 
the special conformal transformation 
has been essentially 
noticed long ago in \cite{fradkin-palchik},  in 
the case of Feynman gauge. It is straightforward to 
check that the measure is invariant under this 
transformation using the BRS formalism 
following reference \cite{fradkin-palchik}.  
In the BRS formalism, the modified term is 
supplied from the field-dependent BRS 
transformation whose Jacobian compensates the violation of 
the special conformal invariance of the gauge fixing term 
$\int d^4 x {1\over 2g_s}{\rm Tr F^2}$. 
Integration over the ghost fields gives the same final 
form for the modified term. 

Because of this modification, the 
conformal Ward identity for the 
effective action $\Gamma$ for the diagonal parts, 
 which is the sum over the diagrams 1PI with respect to the 
diagonal Higgs fields $B$, takes the form
\begin{equation}
\bigl(\delta_K B +4i \langle 
[\triangle_B^{-1}\epsilon \cdot A, X]_{{\rm diagonal}}\rangle\bigr) 
\, {\delta \Gamma \over \delta B}=0 ,
\end{equation}
where the subscript `diagonal' for the commutator indicates 
that only the diagonal part is taken and the bracket 
$\langle \, \cdot \, \rangle$ indicates the expectation value 
with respect to the path integral over the off-diagonal part. 

We can now evaluate the expectation value 
of the additional term. For  small  transverse velocities, 
the lowest nontrivial 
contribution comes from the diagram with one 
vertex 
$$-{2\over g_{{\rm YM}}^2 (2\pi\alpha')^2}
\int d^4 x \, {\rm Tr}\Bigl(
\partial_a B^{\mu} [Y_{\mu}, A_a]\Bigr) 
$$
inserted, which mixes the gauge and the Higgs fields. 
Since this vertex  is already of first order in velocity, 
we can use the static approximation for the rest. 
Then the result for the probe D3-brane in the 
presence of $N$ coincident source D3-branes at the 
origin is 
$$
4i \langle 
[\triangle_B^{-1}\epsilon \cdot A, X]_{{\rm diagonal}}\rangle
=16i g_{{\rm YM}}^2 N
\int {d^4 k\over (2\pi)^4} 
 {1\over (k^2+M^2-i\epsilon)^3}\, \epsilon\cdot \partial B
={g_{{\rm YM}}^2 N \over 2\pi^2 M^2}\, \epsilon\cdot 
\partial B  ,
$$
where $M^2 = \Bigl({r\over 2\pi\alpha'}\Bigr)^2$. 
Note that the background Higgs fields take the following form 
($r^2 = r^{\mu}r_{\mu}$) 
$$
B^{\mu}=
\pmatrix{ 0 & 0 &   \cdots & 0 \cr
                    0 & 0 &  \cdots & 0 \cr 
                     \cdot & \cdot & \cdots & \cdot \cr
                     \cdot & \cdot & \cdots & \cdot \cr
                     0 & 0 &  \cdots &  r^{\mu} \cr} ,
$$
where the only nonzero entry is the last $(N+1,N+1)$ matrix 
element corresponding to the probe D3-brane and the 
$N$ zero diagonal elements represent the 
source D3-branes which can be assumed to be at rest 
in the large $N$ limit. Note also that the 
modification of the conformal transformation 
of the source D3-branes can be neglected 
compared with that of the probe in the large $N$ limit. 

The above result, {\it including the numerical 
coefficient},  precisely gives the 
last term in the transformation law for the 
world-volume ({\it i.e} longitudinal) coordinates of the 
anti de Sitter space-time (\ref{specialadsx}) since 
$R^4/U^2 = (2\pi\alpha')^2g_{{\rm YM}}^2N / 2\pi^2 r^2$. 
The transformation law of the diagonal Higgs fields 
now takes the AdS form (\ref{specialadsu}) 
corresponding to the 
radial coordinate $U$. Remember here that  the 
transformation of the world-volume coordinate and 
the scaling of the fields are oppositely related to each other. 
We have thus succeeded in deriving the 
transformation law in the bulk starting from the 
`boundary' conformal field theory. 
We would like, however, to remind the reader 
that our interpretation is somewhat 
different  as already emphasized before. 
Also, our derivation 
shows that in general the transformation law 
is subject to higher order corrections both 
in velocities and in the Yang-Mills coupling 
constant. Furthermore, the corrections can in principle  
be computed using the conformal Ward identities 
for arbitrary backgrounds.  For general backgrounds, 
however, the modifed transformation rule cannot be 
interpreted in terms of  simple space-time picture 
based on classical geometrical language. 
This  furhter suggests that some kind of collective field 
theory which describes the dynamics of the 
 diagonal part after eliminating the off-diagonal part  
might be a convenient tool 
for establishing the AdS/SYM correspondence in 
more general way.  A discussion 
concerning the relevance of collective 
field theory is given in \cite{jev}. 

As shown in \cite{maldacena}, the modified 
transformation law is very powerful in 
determining the effective action for D3-brane 
on the AdS background 
in the low-velocity approximation. Combined with 
a few assumptions, in particular a  
supersymmetric nonrenormalization theorem,  
the form of the action in the leading approximation in the 
velocity expansion is uniquely 
determined to all {\it classical} orders in the string 
coupling, coinciding with the familiar 
Born-Infeld action.  Thus our result 
implies that the probe D3-brane described 
by the Yang-Mills theory must detect 
the anti de Sitter space-time described by the metric 
(\ref{ads5metric}), since the metric is uniquely 
characterized by the transformation law 
(\ref{specialadsx}) and (\ref{specialadsu}).   

At this juncture, 
let us comment once more on the interpretation 
of the AdS/SYM correspondence. Our result 
indicates that the super Yang-Mills theory in this 
correspondence can indeed 
be interpreted as the effective theory  of D3-branes 
applicable to arbitrary background configurations of 
D3-branes. The AdS 
transformation law emerges upon integrating over the 
off-diagonal degrees of freedom,  which, in the 
picture of $s$-$t$ duality emphasized in the 
beginning of the present paper, represent the 
dynamics of closed strings in the $t$ channel using  the 
dual $s$-channel language.  If there are a large 
number, $N_1$,  of probe D3-branes at a sufficiently 
large radial distance $U$  from the source 
consisting of $N_2 \,(\gg N_1 \gg 1)$ coincident D3-branes, 
the system of the probe 
D-branes is treated as U($N_1$) Yang-Mills theory,  
and the conformal transformation reduces 
to the usual linear one.  It seems that the `boundary' conformal field theory is identified with either 
the probe  or the source Yang-Mills theory with 
reduced gauge group U($N_1$) or U($N_2$), 
respectively.  This suggests that the AdS/SYM correspondence 
may possibly be proven by establishing that the 
gauge-broken part,  U($N_1+N_2$)/U($N_1$)$\times$ 
U($N_2$),  of the whole U($N_1+N_2$) Yang-Mills theory 
describes supergravity in the low-energy 
and the large $N_1, N_2$ limit. 
As is more or less evident from our discussions,  
the low energy (or long-distance) effects on supergravity side 
are in turn related to the high-energy or short-distance 
effects on Yang-Mills side, reflecting the 
`space-time uncertainty' or 'UV/IR' 
relation. This makes the above scenario for 
a possible derivation of the AdS/STM correspondece 
conceptually feasible. 

In reference \cite{jevicki-yoneya}, two of the 
present authors argued that the conformal symmetry 
can be extended to the case of D-particles and may 
be useful for discussing the dynamics of D-particles 
in almost the same sense as for D3-branes.  
The present derivation  of the bulk conformal 
transformation for D3 brane stemmed from our 
investigation along this line. Actually, however, 
the extension of the present formalism to 
other `dilatonic' branes including D-particles 
requires some more intricate treatments and will be 
discussed elsewhere \cite{jky}. 
It is also interesting to see whether  
the present method can be extended 
to other non-dilatonic examples \cite{maldacena} 
of the AdS/CFT correspondence.

\vspace{0.3cm}
The present work grew out of discussions begun at the 
workshop, ``Dualities in String Theory" at ITP, Santa Barbara. 
We would like to thank the organizers of the workshop and 
the stuff of ITP for providing a stimulating 
atmosphere and for hospitality. 
The work of A.J. is supported in part  
by  the Department of Energy under contract DE-FG02-91ER40688-Task A.
The work of Y. K. and T.Y. is supported in part 
by Grant-in-Aid for Scientific  Research (No. 09640337) 
and Grant-in-Aid for International Scientific Research 
(Joint Research, No. 10044061) from the Ministry of  Education, Science and Culture.

\end{document}